\documentclass[english]{article}
\usepackage[T1]{fontenc}
\usepackage[latin9]{luainputenc}
\usepackage{geometry}
\usepackage{amstext}
\usepackage{amssymb}
\usepackage{feyn}
\usepackage{tikz}
\usepackage{graphicx}
\usepackage{array}
\usepackage{booktabs}

\usepackage{amsmath}
\usepackage{amsfonts}
\usepackage{mathrsfs}

\usepackage[normalem]{ulem}
\usepackage{color}
\usepackage{listings,braket}
\usepackage{caption}
\usepackage{subcaption}
\usepackage{float}
\definecolor{darkgreen}{rgb}{0,0.35,0}
\definecolor{Rood}{rgb}{1, 0, 0}

\makeatletter
\usepackage[english]{babel}
\usepackage{feyn}

\makeatother

\begin{document}

\title{\bf BRST invariant scalar and vector operators in the $SU\left(2\right)\times U\left(1\right)$
Higgs model}

{\author{\textbf{D.~Dudal$^{1,2}$}\thanks{ david.dudal@kuleuven.be}, \textbf{D.M.~van Egmond$^3$}\thanks{duivemaria@ictp-saifr.org}, \textbf{G.~Peruzzo$^4$}\thanks{gperuzzofisica@gmail.com},
		\textbf{S.P.~Sorella$^5$}\thanks{silvio.sorella@gmail.com}\\\\\
		\textit{\small $^1$KU Leuven Campus Kortrijk-Kulak, Department of Physics,} \\
		\textit{\small Etienne Sabbelaan 53 bus 7657, 8500 Kortrijk, Belgium}\\
		\textit{\small $^2$Ghent University, Department of Physics and Astronomy, Krijgslaan 281-S9, 9000 Gent, Belgium} \\
		\textit{{\small $^3$ ICTP South American Institute for Fundamental Research Instituto de F\'{i}sica Te\'{o}rica, UNESP - Univ. Estadual Paulista,}}\\
			\textit{{\small	 Rua Dr. Bento Teobaldo Ferraz, 271, 01140-070, S\~ao Paulo, SP, Brazil}}	\\	
		\textit{{\small $^4$Instituto de F\'{i}sica, Universidade Federal Fluminense,
		}}\\
		\textit{{\small Campus da Praia Vermelha, Av. Litor\^{a}nea s/n, }}\\
		\textit{{\small 24210-346, Niter\'{o}i, RJ, Brasil}}\\
		\textit{{\small $^5$UERJ -- Universidade do Estado do Rio de Janeiro,}}\\
		\textit{{\small Instituto de F\'{\i}sica -- Departamento de F\'{\i}sica Te\'orica -- Rua S\~ao Francisco Xavier 524,}}\\
		\textit{{\small 20550-013, Maracan\~a, Rio de Janeiro, Brasil}}\\
	}

\date{}

\maketitle

\begin{abstract}
We present a set of BRST invariant composite operators
in the $SU\left(2\right)\times U\left(1\right)$ Higgs model which exhibit an overlap with the observable scalar and vector particle states of the theory. Some of these operators are non-local in the standard formulation of the Higgs model, however, we show that they can be localized using the Stuckelberg trick, this without changing the physical content. These operators provide thus a  BRST invariant resolution of the physical spectrum of the theory, thereby giving the next step in constructing a manifestly BRST invariant formulation of the particle spectrum of the Electroweak and Standard model.
\end{abstract}

\section{Introduction}

Building on the earlier Abelian case of \cite{Dudal:2019pyg}, in some recent works \cite{Dudal:2020uwb,Dudal:2021dec}, the $SU\left(2\right)$ Higgs model
with a single  scalar field in the fundamental representation, with action
\begin{equation}
	S_{\textrm{Higgs}}^{SU(2)}= \int d^4x \left[ \frac{1}{4}F_{\mu\nu}^aF_{\mu\nu}^a+(\mathscr{D}_{\mu}\varphi)^{\dagger} (\mathscr{D}_{\mu}\varphi) + \frac{\lambda}{2}\left( \varphi^{\dagger}\varphi-\frac{v^2}{2}\right)^2\right],
\end{equation}
has been studied
in the presence of two gauge invariant composite operators,
\begin{eqnarray}
O\left(x\right) & = & \frac{1}{2}\left(h^{2}\left(x\right)+2vh\left(x\right)+\rho^{a}\left(x\right)\rho^{a}\left(x\right)\right)\,,\nonumber \\
R_{\mu}^{a}\left(x\right) & = & -\frac{1}{2}\left[\left(v+h\right)\partial_{\mu}\rho^{a}-\rho^{a}\partial_{\mu}h+\varepsilon^{abc}\rho^{b}\partial_{\mu}\rho^{c}-\frac{g}{2}A_{\mu}^{a}\left(v+h\right)^{2}+g\varepsilon^{abc}A_{\mu}^{b}\rho^{c}\left(v+h\right)+\frac{g}{2}A_{\mu}^{a}\rho^{b}\rho^{b}-gA_{\mu}^{b}\rho^{a}\rho^{b}\right]\,,\nonumber \\
\label{eq:gauge_inv_op_su2}
\end{eqnarray}
where $h\left(x\right)$, $\rho^{a}\left(x\right)$ and $A_{\mu}^{a}\left(x\right)$
represent the Higgs field, the Goldstone field and the gauge field,
respectively. Being directly  related to the Higgs  and
vector fields, $h\left(x\right)$ and $A_{\mu}^{a}\left(x\right)$,
these operators provide a gauge invariant version of the Higgs boson and of the massive vector boson. In fact,
computing the two-point Green's functions of these operator, one gets at tree level
\begin{eqnarray}
\left\langle O\left(k\right)O\left(-k\right)\right\rangle _{\textrm{tree}} & = & v^{2}\left\langle h\left(k\right)h\left(-k\right)\right\rangle _{\textrm{tree}}\,,\nonumber \\
\mathcal{P}_{\mu\nu}\left(k\right)\left\langle R_{\mu}^{a}\left(k\right)R_{\nu}^{b}\left(k\right)\right\rangle _{\textrm{tree}} & = & \frac{v^{2}}{4}\mathcal{P}_{\mu\nu}\left(k\right)\left\langle A_{\mu}^{a}\left(k\right)A_{\mu}^{b}\left(-k\right)\right\rangle _{\textrm{tree}}\,,\label{eq:two_point_su2}
\end{eqnarray}
where $\mathcal{P}_{\mu\nu}\left(k\right)=\delta_{\mu\nu}-\frac{k_{\mu}k_{\nu}}{k^{2}}$
stands for the transverse projector and $v$ is the
classical minimum of the scalar potential $V(\varphi)$.
Eq. (\ref{eq:two_point_su2}) implies that, at the tree level,
$\left\langle O\left(k\right)O\left(-k\right)\right\rangle $ and
$\mathcal{P}_{\mu\nu}\left(k\right)\left\langle R_{\mu}^{a}\left(k\right)R_{\nu}^{b}\left(-k\right)\right\rangle _{\textrm{tree}}$
have the same poles as $\left\langle h\left(-k\right)h\left(-k\right)\right\rangle $
and $\mathcal{P}_{\mu\nu}\left(k\right)\left\langle A_{\mu}^{a}\left(k\right)A_{\nu}^{b}\left(k\right)\right\rangle $,
respectively. In \cite{Dudal:2021dec} we were able to relate $R_{\mu}^{a}\left(x\right)$
with the residual $SU\left(2\right)$ symmetry of the model, which
exists after the spontaneous symmetry breaking and is usually called
\emph{custodial symmetry}. Moreover, it was shown that $R_{\mu}^{a}\left(x\right)$
is in fact the Noether current of this custodial symmetry. This observation has great consequences for the quantum properties of $R_{\mu}^{a}\left(x\right)$,
such as the vanishing of its anomalous dimension, which is expected
for a conserved current. Other major implication is that the longitudinal
component of $\left\langle R_{\mu}^{a}\left(k\right)R_{\nu}^{b}\left(-k\right)\right\rangle $
is exactly momentum independent. Therefore, no propagating mode can be associated
to it, reinforcing the idea that $R_{\mu}^{a}$ is indeed related to a massive vector boson.

The interest in (classically) gauge invariant operators is not only due to the fact
that their BRST invariant quantum extensions lead to Green's functions that are, next to fully renormalizable, also explicitly gauge independent as can be derived from the Slavnov-Taylor identity \cite{Piguet:1984js} \footnote{
In particular they will not depend on the gauge parameter choice, as encoded in the Nielsen identities, see \cite{Nielsen:1975fs,Piguet:1984js,Gambino:1999ai}}, but also because of their spectral
properties. It was shown explicitly up to one-loop \cite{Dudal:2020uwb} that
the K\"all\'{e}n-Lehmann spectral representations of $\left\langle O\left(k\right)O\left(-k\right)\right\rangle $
and $\left\langle R_{\mu}^{a}\left(k\right)R_{\nu}^{b}\left(-k\right)\right\rangle $
render positive definite spectral densities. This important property is not always shared
by the correlation functions of the elementary fields, which in general
are gauge dependent quantities, see \cite{Dudal:2020uwb,Dudal:2019aew,Maas:2020kda}. In addition, these types of operators have the potential to be studied in  lattice gauge theory, allowing for nonperturbative
investigations of the $SU\left(2\right)$ Higgs model, see f.i.~\cite{Jenny:2022atm}. According to
\cite{Fradkin:1978dv}, there should be two distinct regions
in the parameters space of this theory, one Higgs-like and
the other one QCD-like, but these regions are nevertheless analytically connected, so there is no thermodynamical phase transition line separating them. Let us refer to e.g.~\cite{Greensite:2008ss,Greensite:2018mhh,Greensite:2021fyi,Ikeda:2023aid} for more recent discussions, including those using (non-local) order parameters for the custodial symmetry.\\

It should be emphasized that the usual (textbook) treatment of the Higgs mechanism as a spontaneous breaking of local gauge symmetry is ambiguous, since it was already shown in 1975 by Elitzur \cite{Elitzur:1975im} that a local symmetry can never be broken spontaneously. The formulation of the Higgs mechanism with local gauge-invariant operators, was first done by 't Hooft \cite{tHooft:1980xss} and later formalized by Frohlich-Morchio-Strocchi (FMS) \cite{Frohlich:1981yi}, avoids both the notion of spontaneous breaking of a local symmetry as well as complies with the analytical connection between the Higgs and QCD phase. 

As the name itself suggests, in the QCD-like
region the particles are confined as much as in  QCD. Since the positivity
violation of the K\"all\'{e}n-Lehmann spectral representation
has been used as a possible criterion of confinement \cite{Krein:1990sf,Alkofer:2000wg,Cucchieri:2004mf,Bowman:2007du,Fischer:2008uz,Li:2019hyv,Binosi:2019ecz,Dudal:2019gvn,Falcao:2020vyr,Horak:2021syv,Hayashi:2021nnj,Hayashi:2021jju,Siringo:2022dzm,Lechien:2022ieg,Boito:2022rad}, when the  theory goes into
the QCD-like region, positivity violation might be exptected in this region. Therefore, it is important to rule out any possible
effect caused by the gauge dependence, which is another good reason
to look for gauge invariant operators instead of working with elementary, gauge dependent, fields.

From the physical point of view, undoubtly the most interesting case to be considered
is the $SU\left(2\right)\times U\left(1\right)$ Higgs model, an essential ingredient of the Electroweak (Weinberg-Salam) theory and the Standard Model.  This is a more complicated case compared to the $SU(2)$ model, since this is not a simple group and leaves, in the usual language of spontaneous symmetry breaking of the local gauge symmetry, a residual $U(1)$ symmetry at the level of the (gauge-dependent) one-point function for the Higgs field, generated by the electric charge. Moreover, the $SU\left(2\right)\times U\left(1\right)$ theory lacks the custodial symmetry of the pure $SU(2)$ theory which allows, among others things, to derive a set of powerful Ward identities which efficiently controls the renormalization of the BRST-invariant operators. The last difficulty to be faced in the $SU(2)\times U(1)$ case is the reduced number of local BRST-invariant operators with low dimensionality. For instance, \cite{Frohlich:1981yi} presents a set of gauge invariant and gauge covariant fields with the properties mentioned before, however some of these operators have dimension 6.

In this work we attempt at improving this situation with the help of a dressed $U(1)$ BRST-invariant field, $\varphi^h$, much along the same spirit the electron can be dressed to get a truly gauge invariant object, see \cite{Lavelle:1995ty} for a nice review of the Dirac dressing. As the Higgs model is traditionally formulated, such an object can only be non-local. Here, we overcome this obstacle by using an auxiliary Stueckelberg field 	\cite{Stueckelberg:1938hvi,Stueckelberg:1938zz}, $\sigma (x)$, which allows us to extract a local yet equivalent expression for $\varphi^h$. Having this local $U(1)$ BRST invariant dressed field at our disposal, we can employ again some of the operators of the $SU(2)$ model, such as the 't Hooft operators \cite{tHooft:1980xss}, simply by replacing the elementary field $\varphi$ by $\varphi^h$, or better said, its localized version. A similar tactic was employed in e.g.~\cite{Capri:2015ixa,Capri:2016aqq,Capri:2017npq,deBrito:2023qfs}, as it also allows to study the renormalization properties then in terms of local quantum field theory and BRST cohomology \cite{Piguet:1995er}. In this introductory paper, we will not delve into such details and merely introduce the setup for the $SU(2)\times U(1)$ case.

The paper is organized as follows. In Section \ref{A brief review} we review some of the important features of the $SU(2)\times U(1)$ Higgs model, as its classical action, the gauge fixing procedure and the BRST symmetry. In Section \ref{BRST invariant operators} we present a set of BRST-invariant composite operators that have an overlap with one-particle states and may be thought of as the physical  dressed fields of the theory. In Section \ref{conclusions} we present our conclusion
and perspectives.

\section{A brief overview of the $SU\left(2\right)\times U\left(1\right)$ Higgs model}\label{A brief review}

\subsection{The classical action}

To implement the $SU\left(2\right)\times U\left(1\right)$ gauge symmetry in a theory with a complex scalar field in the fundamental representation, \emph{i.e.}
\begin{eqnarray}
\varphi\left(x\right) & = & \left(\begin{array}{c}
\varphi_{1}\left(x\right)\\
\varphi_{2}\left(x\right)
\end{array}\right)\,,
\end{eqnarray}
we choose the Pauli matrices, $\left\{ \frac{\tau^{a}}{2}\,;\,a=1,\,2,\,3\right\} $,
as the generators of the $SU\left(2\right)$ group and the $2\times2$
identity matrix, $I$, as the generator of the $U\left(1\right)$
group. Therefore, we have the following covariant derivative of $\varphi$ :
\begin{eqnarray}
D_{\mu}\varphi & = & \left(I\partial_{\mu}-ig\frac{\tau^{a}}{2}W_{\mu}^{a}-ig'X_{\mu}I\right)\varphi\,,\label{eq:covariant_derivative}
\end{eqnarray}
where $W_{\mu}^{a}$ and $X_{\mu}$ are the gauge fields associated
to $SU\left(2\right)$ and $U\left(1\right)$ groups, respectively, while $g$
and $g'$ in \eqref{eq:covariant_derivative} are the massless gauge
couplings of the theory. We introduce the Euclidean $SU\left(2\right)\times U\left(1\right)$ Higgs action as
\begin{eqnarray}
S_{\textrm{Higgs}} & = & \int d^{4}x\left\{ \frac{1}{4}F_{\mu\nu}^a\left(W\right)F_{\mu\nu}^a\left(W\right)+\frac{1}{4}F_{\mu\nu}\left(X\right)F_{\mu\nu}\left(X\right)\right.\nonumber \\
 &  & \left.+\left(D_{\mu}\varphi\right)^{\dagger}\left(D_{\mu}\varphi\right)+\frac{\lambda}{2}\left(\varphi^{\dagger}\varphi-\frac{v^{2}}{2}\right)^{2}\right\} \,,\label{eq:higgs_action}
\end{eqnarray}
where
\begin{eqnarray}
F_{\mu\nu}^a(W)&=& \partial_{\mu}W^{a}_{\nu}-\partial_{\nu}W^{a}_{\mu}+g\varepsilon^{abc}W_{\mu}^bW_{\nu}^c \, , \nonumber \\
F_{\mu\nu}(X)&=& \partial_{\mu}X_{\nu}-\partial_{\nu}X_{\mu}\, ,
\end{eqnarray}
being $\varepsilon^{abc}$ the Levi-Civita symbol ($\varepsilon^{123}=1$). $\lambda$ is the quartic coupling  and $v$ is a massive parameter related to vacuum expectation value (vev) of $\varphi$.\footnote{We are following here the usual (textbook) formulation of the Higgs mechanism. In reality, Elitzur's theorem forbids a non-vanishing vev for gauge-dependent quantities, and we should speak of a $\textit{minimizing value}$ rather than a vev.} The action \eqref{eq:higgs_action} is invariant under the local $SU(2)\times U(1)$ gauge
transformations
\begin{eqnarray}
\varphi' & = & e^{-ig\beta^{a}\frac{\tau^{a}}{2}-ig'\alpha I}\varphi\,,\nonumber \\
\varphi'^{\dagger} & = & \varphi^{\dagger}e^{ig\beta^{a}\frac{\tau^{a}}{2}+ig'\alpha I}\,,\nonumber \\
X'_{\mu} & = & X_{\mu}-\partial_{\mu}\alpha\,,\nonumber \\
{W'}_{\mu}^{a}\frac{\tau^{a}}{2} & = & e^{-ig\frac{\tau^{a}}{2}\beta^{a}}W_{\mu}^{b}\frac{\tau^{b}}{2}e^{ig\frac{\tau^{c}}{2}\beta^{c}}-\frac{\tau^{a}}{2}\partial_{\mu}\beta^{a}\,.\label{eq:gauge_transformations}
\end{eqnarray}
Let us say more about the massive parameter $v$ in \eqref{eq:higgs_action}. As already mentioned, $v$ is related to vev of the scalar field. Analyzing the classical potential resulting from $S_{\textrm{Higgs}}$,
which is
\begin{eqnarray}
V\left(\varphi\right) & = & \frac{\lambda}{2}\left(\varphi^{\dagger}\varphi-\frac{v^{2}}{2}\right)^{2}\,,\label{eq:Higgs_potential}
\end{eqnarray}
one gets that its minimums occurs when $\varphi^{\dagger}\varphi=\frac{v^{2}}{2}$.
This result implies that $\left\langle \varphi\right\rangle \neq0$,
which indicates the spontaneous symmetry breaking of the $SU\left(2\right)\times U\left(1\right)$
symmetry. We choose to expand around the classical minimum of $V\left(\varphi\right)$ such that
\begin{eqnarray}
\left\langle \varphi\right\rangle  & = & \frac{v}{\sqrt{2}}\left(\begin{array}{c}
1\\
0
\end{array}\right),\label{eq:phi_expectation_value}
\end{eqnarray}
then, all of the original generators of the group are broken, namely,
\begin{equation}
	\frac{\tau^a}{2} \left\langle \varphi\right\rangle \neq 0, \;\;\;\;\; \frac{I}{2} \left\langle \varphi\right\rangle \neq 0. 
 \label{eq:generators_broken}
\end{equation}
However, changing to any other basis which contains
\begin{eqnarray}
Q & = & \tau^{3}-I
\end{eqnarray}
as one of the generators, it is straightforward to see that the vacuum
is invariant with respect to the $U\left(1\right)$ group generated
by $Q$. As usual, we identify $Q$ with the electric charge operator. According to the pattern of the symmetry breaking,  there must be three massive gauge bosons, which will
be identified with the $W^{+}$, $W^{-}$ and $Z^{0}$, and one massless
gauge boson, the photon $\gamma$.

To obtain the correct propagating fields, let us parametrize $\varphi$ around the minimum which leads to (\ref{eq:phi_expectation_value}),
\emph{i.e.}
\begin{eqnarray}
\varphi\left(x\right) & = & \frac{1}{\sqrt{2}}\left(\left(v+h\left(x\right)\right)I+i\tau^{a}\rho^{a}\left(x\right)\right)\left(\begin{array}{c}
1\\
0
\end{array}\right)\,,\label{eq:field_para}
\end{eqnarray}
where $h\left(x\right)$ and $\rho^{a}\left(x\right)$ stand for the
Higgs field and the would be Goldstone fields, respectively. Rewriting
the Higgs action in terms of (\ref{eq:field_para}), one gets
\begin{eqnarray}
S_{\textrm{Higgs}} & = & \int d^{4}x\left\{ \frac{1}{4}F_{\mu\nu}^a(W)F_{\mu\nu }^a(W)+\frac{1}{4}F_{\mu\nu}(X)F_{\mu\nu}(X)\right.\nonumber \\
  &  & +\frac{1}{2}\left(\partial_{\mu}h\right)^{2}+\frac{1}{2}\left(\partial_{\mu}\rho^{a}\right)^{2}\nonumber \\
 & &-\frac{g}{2}W_{\mu}^{a}\left[\left(v+h\right)\left(\partial_{\mu}\rho^{a}\right)-\left(\partial_{\mu}h\right)\rho^{a}+\varepsilon^{abc}\left(\partial_{\mu}\rho^{b}\right)\rho^{c}\right]\nonumber \\
 &  & +\frac{g^{2}}{8}W_{\mu}^{a}W_{\mu}^{a}\left(\left(v+h\right)^{2}+\rho^{b}\rho^{b}\right)\nonumber \\
 &  &-\frac{g'}{2}X_{\mu}\left[\left(\partial_{\mu}\rho^{3}\right)\left(v+h\right)-\left(\partial_{\mu}h\right)\rho^{3}-\left(\partial_{\mu}\rho^{a}\right)\rho^{b}\varepsilon^{ab3}\right]\nonumber \\
 &  & +\frac{g'^{2}}{8}X_{\mu}X_{\mu}\left(\left(v+h\right)^{2}+\rho^{a}\rho^{a}\right)\nonumber \\
 &  & +\frac{gg'}{4}W_{\mu}^{3}X_{\mu}\left(v+h\right)^{2}-\frac{gg'}{4}W_{\mu}^{a}X_{\mu}\rho^{b}\left(v+h\right)\varepsilon^{ab3}\nonumber \\
 &  & +\frac{gg'}{2}W_{\mu}^{a}X_{\mu}\rho^{a}\rho^{3}-\frac{gg'}{4}W_{\mu}^{3}X_{\mu}\rho^{a}\rho^{a}\nonumber \\
 &  & \left.+\frac{\lambda}{8}\left[h^{4}+4v^{2}h^{2}+\left(\rho^{a}\rho^{a}\right)^{2}+4vh^{3}+2h^{2}\rho^{a}\rho^{a}+4vh\rho^{a}\rho^{a}\right]\right\} \,.\label{eq:higgs_action_rewritten}
\end{eqnarray}
Looking at eq.(\ref{eq:higgs_action_rewritten}), one sees that $h$ acquired a mass
\begin{equation}
m_h=\sqrt{\lambda} v,
\end{equation}
whereas $\rho^a$ is a massless field, consistent with it being a Goldstone field. Furthermore, there is a massive term involving the gauge fields, which is
\begin{eqnarray}
\frac{g^{2}v^{2}}{8}W_{\mu}^{3}W_{\mu}^{3}+\frac{g'^{2}v^{2}}{8}X_{\mu}X_{\mu}+\frac{gg'v^{2}}{4}W_{\mu}^{3}X_{\mu} & = & \frac{v^{2}}{8}\left(gW_{\mu}^{3}+g'X_{\mu}\right)^{2}\,.\label{eq:gauge_massive_term}
\end{eqnarray}
Proceeding as in \cite{Weinberg:1967tq}, one can introduce a set of new fields, $\left\{ Z_{\mu},\,A_{\mu}\right\} $,
\begin{equation}
Z_{\mu} = \sin\theta_{W}W_{\mu}^{3}+\cos\theta_{W}X_{\mu}\,,\qquad
A_{\mu} = \cos\theta_{W}W_{\mu}^{3}-\sin\theta_{W}X_{\mu}\,,
\end{equation}
where the Weinberg angle, $\theta_{W}$, is defined by
\begin{equation}
\sin\theta_{W}  =  \frac{g}{\sqrt{g^{2}+g'^{2}}}\,,\qquad
\cos\theta_{W}  = \frac{g'}{\sqrt{g^{2}+g'^{2}}}\,.
\end{equation}
Therefore, we get
\begin{eqnarray}
\frac{g^{2}v^{2}}{8}W_{\mu}^{3}W_{\mu}^{3}+\frac{g'^{2}v^{2}}{8}X_{\mu}X_{\mu}+\frac{gg'v^{2}}{4}W_{\mu}^{3}X_{\mu} & = & \frac{v^{2}g^{2}}{8\sin^{2}\theta_{W}}Z_{\mu}^{2}\,,
\end{eqnarray}
which means that $Z_{\mu}$ picks up a mass
\begin{eqnarray}
m_{Z}^{2} & = & \frac{v^{2}g^{2}}{4\sin^{2}\theta_{W}}\,,
\end{eqnarray}
whereas $A_{\mu}$ is a massless field. For the others gauge bosons, $W^{1}$
and $W^{2}$, their mass reads
\begin{eqnarray}
m_{W^{1}}^{2} & = & m_{W^{2}}^{2}=\frac{v^{2}g^{2}}{4}\,.
\end{eqnarray}
Instead of working with $W^{1}$ and $W^{2}$, we can, like the other gauge fields, change the basis and use the charged
fields
\begin{eqnarray}
W^{+} & = & W_{1}+iW_{2}\,,\qquad W^{-} =  W_{1}-iW_{2}. \label{eq:bosons_W}
\end{eqnarray}

\subsection{BRST symmetry and gauge fixing}
To quantize the Higgs action, it is necessary to fix the gauge. For this, we follow the BRST quantization method. According
to (\ref{eq:gauge_transformations}), for infinitesimal gauge transformations,
we get the following infinitesimal transformations for  the Higgs and
the would be Goldstone field,
\begin{eqnarray}
\delta h & = & \frac{g}{2}\beta^{3}\rho^{3}+\frac{g}{2}\beta^{\alpha}\rho^{\alpha}+g'\alpha\rho^{3}\,,\nonumber \\
\delta\rho^{3} & = & -g'\alpha\left(v+h\right)-\frac{g}{2}\beta^{3}\left(v+h\right)+\frac{g}{2}\varepsilon^{\alpha\beta}\beta^{\alpha}\rho^{\beta}\,,\nonumber \\
\delta\rho^{\alpha} & = & g'\alpha\varepsilon^{\alpha\beta}\rho^{\beta}-\frac{g}{2}\beta^{\alpha}\left(v+h\right)+\frac{g}{2}\varepsilon^{\alpha\beta}\left(\beta^{\beta}\rho^{3}-\beta^{3}\rho^{\beta}\right)\,,\label{eq:inf_gauge_transf_higgs_goldstone}
\end{eqnarray}
where $\varepsilon^{\alpha\beta3}=\varepsilon^{\alpha\beta}$. Therefore,
the BRST transformations are defined as
\begin{eqnarray}
sX_{\mu} & = & -\partial_{\mu}c\,,\nonumber \\
sW_{\mu}^{3} & = & -\partial_{\mu}c^{3}+g\varepsilon^{\alpha\beta}W_{\mu}^{\alpha}c^{\beta}\,,\nonumber \\
sW_{\mu}^{\alpha} & = & -\partial_{\mu}c^{\alpha}+g\varepsilon^{\alpha\beta}\left(W_{\mu}^{3}c^{\beta}-W_{\mu}^{\beta}c^{3}\right)\,,\nonumber \\
sh & = & \frac{g}{2}c^{3}\rho^{3}+\frac{g}{2}c^{\alpha}\rho^{\alpha}+g'c \rho^{3}\,,\nonumber \\
s\rho^{3} & = & -g'c\left(v+h\right)-\frac{g}{2}c^{3}\left(v+h\right)+\frac{g}{2}\varepsilon^{\alpha\beta}c^{\alpha}\rho^{\beta}\,,\nonumber \\
s\rho^{\alpha} & = & g'\varepsilon^{\alpha\beta}c\rho^{\beta}-\frac{g}{2}c^{\alpha}\left(v+h\right)+\frac{g}{2}\varepsilon^{\alpha\beta}\left(c^{\beta}\rho^{3}-c^{3}\rho^{\beta}\right)\,,\nonumber \\
sc & = & 0\,,\nonumber \\
s\overline{c} & = & ib\,,\qquad sb=0\,,\nonumber \\
sc^{3} & = & \frac{g}{2}\varepsilon^{\alpha\beta}c^{\alpha}c^{\beta}\,,\nonumber \\
sc^{\alpha} & = & g\varepsilon^{\alpha\beta}c^{\beta}c^{3}\,,\nonumber \\
s\overline{c}^{3} & = & ib^{3}\,, \qquad sb^3=0\,,\nonumber \\
s\overline{c}^{\alpha} & = & ib^{\alpha}\,,\qquad sb^{\alpha}=0\, ,\label{eq:BRST_transf}
\end{eqnarray}
where $s$ is the nilpotent BRST operator, \emph{i.e.} $s^{2}=0$, and
\begin{eqnarray}
\left(c\,,\overline{c}\,,\,c^{3},\,\overline{c}^{3},\,c^{\alpha},\,\overline{c}^{\alpha}\right)
\end{eqnarray}
are the ghost and antighost fields, and
\begin{eqnarray}
\left(b\,,\,b^{3},\,b^{\alpha}\right)
\end{eqnarray}
are the auxiliary Nakanishi-Lautrup fields.

In the Higgs model, due to possible infrared singularities, see e.g.~\cite{Aitchison:1983ns}, not many interesting gauge options are available. The most commonly used one  is the
$R_{\xi}$-gauge \cite{tHooft:1971qjg, Fujikawa:1972fe} and characterized by the gauge parameter $\xi$. As we are interested in correlation functions of BRST invariant operators, therefore gauge independent quantities, we take the particular case $\xi=0$ which corresponds to the Landau gauge.  Then the gauge fixing term is
\begin{eqnarray}
S_{\textrm{gf}} & = & s\int d^{4}x\left(\overline{c}\partial_{\mu}X_{\mu}+\overline{c}^{3}\partial_{\mu}W_{\mu}^{3}+\overline{c}^{\alpha}\partial_{\mu}W_{\mu}^{\alpha}\right) \nonumber \\
 & = & \int d^{4}x\left[ib\partial_{\mu}X_{\mu}+\overline{c}\partial^{2}c+ib^{3}\partial_{\mu}W_{\mu}^{3}-\overline{c}^{3}\partial_{\mu}\left(-\partial_{\mu}c^{3}+g\varepsilon^{\alpha\beta}W_{\mu}^{\alpha}c^{\beta}\right)\right. \nonumber \\
 &  & \left.+ib^{\alpha}\partial_{\mu}W_{\mu}^{\alpha}-\overline{c}^{\alpha}\partial_{\mu}\left(-\partial_{\mu}c^{\alpha}+g\varepsilon^{\alpha\beta}\left(W_{\mu}^{3}c^{\beta}-W_{\mu}^{\beta}c^{3}\right)\right)\right].
\end{eqnarray}
Due to the nilpotency of the BRST operator, $S_{\textrm{gf}}$ is BRST invariant by construction. Consequently, the starting action
\begin{equation}
S=S_{\textrm{Higgs}}+S_{\textrm{gf}}, \label{starting_action}
\end{equation}
is BRST invariant too:
\begin{equation}
s S =0 \;. \label{invBRSTS}
\end{equation}

We are now ready to introduce a set of composite gauge invariant operators associated to the non-invariant elementary fields $(\gamma, Z^0, W^+, W^-,h)$.

\section{BRST invariant operators}\label{BRST invariant operators}

As emphasized in the Introduction, the aim is to construct  a set of BRST invariant
operators which  have an overlap with the elementary particle states of the
$SU\left(2\right)\times U\left(1\right)$ Higgs model. Moreover, we
select the operators which have the lowest possible dimension to describe the vector and scalar excitations. We introduce and discuss such operators in the following subsections.

\subsection{Scalar sector}\label{subsec:scalar_operator}
The Higgs particle is a scalar chargeless particle and can be associated with the BRST invariant operator $\varphi^{\dagger} \varphi$. However this operator is reducible to a sum of two BRST invariant quantities,
\begin{eqnarray}
O(x) & = & \varphi^{*}(x)\varphi (x)-\frac{v^2}{2} = \frac{1}{2}(h^2(x)+2vh(x)+\rho^a\rho^a(x))
\end{eqnarray}
and
\begin{eqnarray}
\frac{v^2}{2}.
\end{eqnarray}
Such operators are always present in any Higgs model, so no new operators are introduced here. As the non-linear terms do not contribute at the tree-level, we get
\begin{equation}
\langle O\left(x\right) O\left(y\right) \rangle_{\textrm{tree}}=v^2\langle h\left(x\right) h\left(y\right) \rangle_{\textrm{tree}}, \label{eq:scalar_operator}
\end{equation}
which means that $O\left(x\right)$ has an overlap with the Higgs particle state. Let us refer to \cite{Dudal:2019pyg,Dudal:2021pvw,Dudal:2021dec} for more about this operator and its properties in the $U(1)$ or $SU(2)$ case.

\subsection{Vector sector}

\subsubsection{$SU(2)$ case}

In the case of $SU\left(2\right)$ Higgs model, to describe the three vector
bosons, the gauge-invariant operators presented by 't Hooft \cite{tHooft:1980xss} are
\begin{eqnarray}
O_{\mu}^{+}\left(\varphi\right) & = & \varphi^{T}\left(\begin{array}{cc}
0 & 1\\
-1 & 0
\end{array}\right)\mathscr{D}_{\mu}\varphi\,,\nonumber \\
O_{\mu}^{-}\left(\varphi\right) & = & \left(O_{\mu}^{+}\right)^{\ast}\,,\nonumber \\
O_{\mu}^{3}\left(\varphi\right) & = & i\varphi^{\dagger}\mathscr{D}_{\mu}\varphi\,,\label{eq:thooft_operators}
\end{eqnarray}
where,
\begin{equation}
\mathscr{D}_{\mu}=\partial_{\mu}-\frac{ig}{2}\tau^{a}W_{\mu}^{a} \label{covariant_der_su2}
\end{equation}
is the $SU(2)$ covariant derivative. These operators are
not only BRST invariant but also (classically) gauge invariant, as a consequence of their independence from the Faddeev-Popov ghosts. As we will show in the next section, these operators are the gauge-invariant extensions of the charged gauge fields $W^{+/-}$ and the neutral gauge field $Z$.  they can also be recombined as in eq.\eqref{eq:gauge_inv_op_su2}, namely:
\begin{eqnarray}
R_{\mu}^{1} & = & \frac{i}{2}\left(O_{\mu}^{+}-O_{\mu}^{-}\right)\,,\nonumber \\
R_{\mu}^{2} & = & \frac{1}{2}\left(O_{\mu}^{+}+O_{\mu}^{-}\right)\,,\nonumber \\
R_{\mu}^{3} & = & O_{\mu}^{3}-\frac{i}{2}\partial_{\mu}O\,,\label{eq:R_thoft}
\end{eqnarray}
This particular form \eqref{eq:R_thoft} is not only notationally convenient but also because these operators  $R_{\mu}^{a}$ transform as  vectors
in the adjoint representation with respect to the \emph{custodial transformations},
see \cite{Dudal:2021dec}. Indeed, taking the linear part of $R_{\mu}^{a}$,
one gets
\begin{eqnarray}
\left.R_{\mu}^{a} (\varphi)\right|_{\textrm{linear}} & = & \frac{gv^{2}}{4}A_{\mu}^{a}-\frac{v}{2}\partial_{\mu}\rho^{a}\,.\label{eq:R_linear}
\end{eqnarray}
Looking at \eqref{eq:R_linear}, one can already glimpse that $R_{\mu}^{a}(\varphi)$
displays the precise combination between $A_{\mu}^{a}$ and $\rho^{a}$
that leads to a  massive vector field. Since $a$
runs from $1$ to $3$, there are three massive vector fields and, consequently, three massive gauge bosons.

\subsubsection{$SU(2)\times U(1)$ case}

We would like to find BRST invariant objects similar to \eqref{eq:R_thoft}
in the $SU\left(2\right)\times U\left(1\right)$ Higgs model.

Let us first focus on the $SU(2)$ part. Unfortunately,
as it is straightforward to check, the operators in \eqref{eq:thooft_operators}
are not invariant under the whole $SU\left(2\right)\times U\left(1\right)$
group. This lack of invariance is caused by $\varphi$, which is not
invariant with respect to the $U\left(1\right)$ transformation generated by the hypercharge $I/2$. So,
if instead of $\varphi$ in \eqref{eq:thooft_operators}, we introduce the
dressed $U\left(1\right)$ gauge invariant field
\begin{eqnarray}
\varphi^{h}\left(x\right) & = & e^{-ig'\int d^{4}y\,\left(\frac{1}{\partial^{2}}\right)_{x,y}\left(\partial_{\mu}X_{\mu}\right)\left(y\right)}\varphi(x),\label{eq:phi_h_nonlocal}
\end{eqnarray}
then the composite operators
\begin{eqnarray}
O_{\mu}^{+}\left(\varphi^{h}\right),\,O_{\mu}^{-}\left(\varphi^{h}\right),\,O_{\mu}^{3}\left(\varphi^{h}\right),\label{eq:thooft_op_su2_u1}
\end{eqnarray}
are fully gauge invariant as it can be easily checked. Notice that the same substitution $\varphi\to\varphi^h$ has no effect at the level of the already introduced scalar operator \eqref{eq:scalar_operator}.

Obviously, $\varphi^{h}\left(x\right)$
is not particularly useful in the form \eqref{eq:thooft_op_su2_u1},
its non-locality is the first thing that we need to overcome\footnote{We draw attention here to the special role played the Landau gauge $\partial_\mu X_\mu=0$, in which case $\varphi^h$ effectively reduces to $\varphi$. Using a similar analysis as in \cite{Capri:2018ijg}, it can then be shown that gauge invariant correlation functions containing $\varphi^h$ can be computed in the Landau gauge whilst replacing $\varphi^h$ with $\varphi$. }. This step
can be treated by introducing an auxiliary field $\sigma\left(x\right)$,
a Stueckelberg-type field, satisfying the constraint
\begin{eqnarray}
\partial^{2}\sigma & = & \partial_{\mu}X_{\mu}.\label{eq:constraint_stueckelberg}
\end{eqnarray}
Then,  eq.\eqref{eq:phi_h_nonlocal} can be rewritten in a local way, {\it i.e}
\begin{eqnarray}
\varphi^{h}\left(x\right) & = & e^{-ig'\sigma\left(x\right)}\varphi(x).\label{eq:phi_h_local}
\end{eqnarray}
Furthermore, the operators in \eqref{eq:thooft_op_su2_u1} maintain the (nilpotent) BRST invariance
by demanding that $\sigma$ transforms as
\begin{eqnarray}
s\sigma & = & -c.\label{eq:stueckelberg_brs_transf}
\end{eqnarray}
In order to have a consistent set of local BRST invariant operators, it remains to implement the constraint \eqref{eq:constraint_stueckelberg}, a task easily done by means of the functional integral. In fact, with  the introduction of the Stueckelberg field $\sigma$, any functional $F$ of $\varphi^h$ in its non-local form can be rewritten as
\begin{eqnarray}
F[\varphi^h_{\textrm{non-local}}]&=& \int D \sigma \det (-\partial^2) \delta (\partial^2\sigma-\partial_{\mu}X_{\mu})F[\varphi^h_{\textrm{local}}]\,, \label{eq:functional_phih_nonlocal}
\end{eqnarray}
where the condition \eqref{eq:constraint_stueckelberg} is imposed by the Dirac delta function. The $\det (-\partial^2)$ in \eqref{eq:functional_phih_nonlocal} is necessary to properly take into account the argument of the delta function. The determinant and the delta function can be exponentiated by introducing extra ghosts and auxiliary fields, $(\eta\,,\bar{\eta})$ and $\tau$, respectively, resulting in
\begin{eqnarray}
F[\varphi^h_{\textrm{non-local}}]= \int D \sigma D\eta D\bar{\eta}\,e^{-\int d^4x\,\left[i\tau (\partial^2\sigma -\partial_{\mu}X_{\mu})+\bar{\eta}\partial^2 \eta \right]}F[\varphi^h_{\textrm{local}}]\,. \label{eq:functional_phih_local}
\end{eqnarray}
Naturally, the argument of the exponential in \eqref{eq:functional_phih_local} can be considered part of the starting action of an extended theory with the (BRST singlet) extra fields $(\eta,\, \bar{\eta},\,\tau)$, namely,
\begin{eqnarray}
S &=& S_{\textrm{Higgs}}+S_{\textrm{gf}}+\int d^4x\,\left[i\tau (\partial^2\sigma -\partial_{\mu}X_{\mu})+\bar{\eta}\partial^2 \eta \right]. \label{action_local_stueckelberg}
\end{eqnarray}
Let us write down explicitly the  expressions of the local operators $(O^3_\mu, O^+_\mu, O^-_\mu)$ in terms of the fundamental fields we are working with:
\begin{eqnarray}
O_{\mu}^{3}\left(\varphi^h\right) & = & -\frac{ig'}{2}\left(\partial_{\mu}\sigma\right)\left[\left(v+h\right)^{2}+\rho^{a}\rho^{a}\right]\nonumber \\
&  & +\frac{1}{2}\partial_{\mu}O-\frac{i}{2}\left(\rho^{3}\partial_{\mu}h-\left(v+h\right)\partial_{\mu}\rho^{3}+\varepsilon^{ab3}\partial_{\mu}\rho^{a}\rho^{b}\right)\nonumber \\
&  & -\frac{ig}{4}\left[\left(v+h\right)^{2}W_{\mu}^{3}+2\left(v+h\right)W_{\mu}^{a}\rho^{b}\varepsilon^{abc}+2W_{\mu}^{a}\rho^{3}\rho^{a}-W_{\mu}^{3}\rho^{b}\rho^{b}\right]\,,
\end{eqnarray}
\begin{eqnarray}
O_{\mu}^{+}\left(\varphi^h\right) & = & -\frac{ig'}{2}e^{-2ig'\sigma}\left(\partial_{\mu}\sigma\right)\left(v+h\right)i\left(\rho^{1}+i\rho^{2}\right)-\frac{ig'}{2}e^{-2ig'\sigma}\left(\partial_{\mu}\sigma\right)i\left(v+h\right)\left(-\rho^{1}-i\rho^{2}\right)\nonumber \\
&  & +\frac{1}{2}e^{-2ig'\sigma}\left(v+h\right)i\left(\partial_{\mu}\rho^{1}+i\partial_{\mu}\rho^{2}\right)+\frac{1}{2}e^{-2ig'\sigma}i\left(-\rho^{1}-i\rho^{2}\right)\left(\partial_{\mu}h\right)\nonumber \\
&  & -\frac{1}{2}e^{-2ig'\sigma}\left(-\rho^{1}\partial_{\mu}\rho^{3}+\rho^{3}\partial_{\mu}\rho^{1}-i\rho^{2}\partial_{\mu}\rho^{3}+i\rho^{3}\partial_{\mu}\rho^{2}\right)\nonumber \\
&  & -\frac{ig}{4}e^{-2ig'\sigma}\left(v+h\right)^{2}\left(W_{\mu}^{1}+iW_{\mu}^{2}\right)+\frac{ig}{4}e^{-2ig'\sigma}\left(v+h\right)\left(W_{\mu}^{2}\rho^{3}-W_{\mu}^{3}\rho^{2}+i\varepsilon^{2ab}W_{\mu}^{a}\rho^{b}\right)\nonumber \\
&  & +\frac{g}{4}e^{-2ig'\sigma}\left(v+h\right)\left(-W_{\mu}^{3}\rho^{1}+W_{\mu}^{1}\rho^{3}-iW_{\mu}^{3}\rho^{2}+iW_{\mu}^{2}\rho^{3}\right)\nonumber \\
&  & +\frac{i}{4}e^{-2ig'\sigma}g\left[-\left(W_{\mu}^{1}\rho^{1}\rho^{1}+iW_{\mu}^{1}\rho^{2}\rho^{1}\right)-i\left(W_{\mu}^{1}\rho^{1}\rho^{2}+iW_{\mu}^{1}\rho^{2}\rho^{2}\right)\right.\nonumber \\
&  & +i\left(W_{\mu}^{2}\rho^{1}\rho^{1}+iW_{\mu}^{2}\rho^{3}\rho^{1}\right)-\left(W_{\mu}^{2}\rho^{1}\rho^{2}+iW_{\mu}^{2}\rho^{2}\rho^{2}\right)-\left(W_{\mu}^{3}\rho^{1}\rho^{3}+iW_{\mu}^{3}\rho^{2}\rho^{3}\right)\nonumber \\
&  & \left.-iW_{\mu}^{1}\rho^{3}\rho^{3}+W_{\mu}^{2}\rho^{3}\rho^{3}-iW_{\mu}^{3}\rho^{3}\rho^{1}-W_{\mu}^{3}\rho^{3}\rho^{2}\right]\,.
\end{eqnarray}
To obtain $O^{-}_{\mu}(\varphi)$ we can  take the complex conjugate of  $O^{+}_{\mu}(\varphi)$, according to Eq. \eqref{eq:thooft_operators}.  Analyzing $O_{\mu}^{3}(\varphi^h)$ we see that it reduces to a sum of two BRST invariant operators, one of them is the scalar operator $O(x)$ introduced in Subsection \ref{subsec:scalar_operator}. This feature, see also \cite{Dudal:2019pyg,Dudal:2021dec}, suggests to work with
\begin{equation}
\widetilde{O}_{\mu}^{3} =  i \left( O^{3}_{\mu}-\frac{1}{2}\partial_{\mu}O\right) \label{O_tilde_def}
\end{equation}
instead of  $O_{\mu}^{3}$. Taking the linear part of the vector operators, one gets
\begin{eqnarray}
\left.\widetilde{O}_{\mu}^{3}(\varphi^h)\right|_{\textrm{linear}} &  = & \frac{v^{2}g'}{2}\partial_{\mu}\sigma-\frac{v}{2}\partial_{\mu}\rho^{3}+\frac{gv^{2}}{4}W_{\mu}^{3}\, , \nonumber \\
\left.O_{\mu}^{\pm}(\varphi^h)\right|_{\textrm{linear}} & = & \mp\frac{igv^{2}}{4}\left(W_{\mu}^{1}\pm iW_{\mu}^{2}\right)\pm\frac{1}{2}iv\left(\partial_{\mu}\rho^{1}\pm i\partial_{\mu}\rho^{2}\right) \,. \label{eq:O_vec_linear}
\end{eqnarray}
Remembering the definition \eqref{eq:bosons_W} of the $W^{\pm}_\mu$ field, we see that $O_{\mu}^{\pm}$ can indeed be interpreted as its BRST invariant version. We do not need to worry about the presence of the Goldstone field since it is well-known that it does not represent any physical mode, more precisely it belongs to the trivial part of the BRST cohomology. Analogously, $\widetilde{O}_{\mu}^{3}$ is associated with $W^3_{\mu}$, here also taking into account that $\sigma$ and $c$ have a BRST doublet type structure\footnote{Strictly speaking, $c$ and $\sigma$ are not a doublet, as $c$ also appears in the BRST transformations of other fields, see \eqref{eq:BRST_transf}.}.

We observe that, as expected, the operator $R_\mu^{a}\left(\varphi^h \right)$ is now not conserved on-shell. If it were it would have the meaning of a conserved current and, as such, it would generate a global symmetry of the action via Noether's theorem.

 To finalize the construction of gauge invariant operators, we turn to the gauge boson of the Abelian $U(1)$ sector whic can be described in several ways. The simplest possible way is just to use the BRST invariant operator
\begin{eqnarray}
\partial_{\nu}F_{\nu\mu}(X)&=& \partial^2 X_{\mu}-\partial_{\mu}\partial_{\nu}X_{\nu}, \label{eq:transverse_x}
\end{eqnarray}
which is essentially the transverse part of $X_{\mu}$. This type of operator always exists in Abelian theories, as the QED and Abelian Higgs model. In addition to \eqref{eq:transverse_x}, another vector operator with dimension 3, and with interesting properties, is
\begin{eqnarray}
V_{\mu} &=& -\frac{1}{2}\left[ \partial_{\mu} \rho^3 (v+h)-\rho^3 \partial_{\mu} h + \varepsilon^{\alpha \beta}\rho^{\alpha} \partial_{\mu}\rho^{\beta}  \right] +\frac{g'}{4}X_{\mu}(v^2+2vh+h^2+\rho^a\rho^a) \nonumber \\
& & +\frac{g }{4}\left[ W_{\mu}^3(v+h)^2-\varepsilon^{\alpha\beta}W_{\mu}^a\rho^b(v+h)+2W_{\mu}^a\rho^a\rho^3-W_{\mu}^3\rho^a\rho^a \right] \, .\label{V_operator}
\end{eqnarray}
The easiest way to see the BRST invariance of $V_{\mu}$ is to look at the equations of motion of $X_{\mu}$, which is
\begin{eqnarray}
\frac{\delta S_{\textrm{Higgs}}}{\delta X_{\mu} } &=& -\partial_{\nu} F_{\nu\mu}(X)+g'V_{\mu}\, . \label{eq:equation_motion_X}
\end{eqnarray}
Since $\left[s,\frac{\delta }{\delta X_{\mu}}\right]=0$ and $sS_{\textrm{Higgs}}=sF_{\mu\nu}(X)=0$, it follows immediately that
\begin{equation}
s V_{\mu}=0\,. \label{eq:brst_V}
\end{equation}
The BRST invariance of $V_{\mu}$ can also be checked by observing that it can be rewritten as
\begin{equation}
V_{\mu}=ig'\left[\varphi^{\dagger}D_{\mu}\varphi-(D_{\mu}\varphi)^\dagger \varphi \right],
\end{equation}
which is manifestly a gauge invariant operator. Again, taking the linear part of $V_{\mu}$, we get
\begin{eqnarray}
V_{\mu}|_{\textrm{linear}}&=&-\frac{v}{2}\partial_{\mu}\rho^3+\frac{g'v^2}{4}X_{\mu}+\frac{gv^2}{4}W_{\mu}^3.
\end{eqnarray}
This operator is a bit different than the others, here the association with an elementary field is not immediate. However, we can combine $\tilde{O}_{\mu}^3$, see eq.\eqref{eq:O_vec_linear}, with $V_{\mu}$ to obtain a BRST-invariant operator which has in the linear part only the gauge field $X_{\mu}$.

\section{Conclusions} \label{conclusions}
When we consider the physics of a gauge theory, we generally deal with gauge dependent quantities, under the form of correlation functions of elementary fields. To ensure that physical quantities are gauge independent, we often rely on the sophisticated mechanisms grounded in the BRST symmetry, as explicitly expressed by the Slavnov-Taylor identity and the Nielsen identities. According to these identities, the direct way to obtain gauge independent quantities is by working with BRST invariant operators, the natural quantum generalization of classically gauge invariant field combinations.

In general, these BRST invariant operators are composite. This does neither represent a technical limitation, see \cite{Haag:1992hx}, nor does it obscure the physical understanding in the case of the Higgs model. The operators we presented in this work are indeed composite, but the particle states that have an overlap with them also overlap with some of the elementary fields, so we can still say without reservation that they are associated to fundamental and not composite particles, at least in the weak coupling limit. Notice that in this work, we have made use of two different methods of introducing composite operators in order to achieve gauge-invariance: composite operators following 't Hooft \cite{tHooft:1980xss} and composite operators following the Stueckelberg formulations. Although these methods bear similarities, an important difference is that in the first method, elementary degrees of freedom are combined into composite states, and can be regarded as bound states in the same fashion as e.g. hadron operators in QCD \cite{Maas:2017wzi}. In the second case, auxiliary (unphysical) fields are introduced to guarantee the gauge-invariance of the construction, with no implications for the physical spectrum. 

As the $SU(2)\times U(1)$ Higgs model is an essential building block of the Standard Model, it can be used as a less complicated but still quite interesting laboratory to scrutinize, thinking about items such as spontaneous symmetry breaking, triviality, and asymptotic freedom. 

A first step in continuing the research set out in this work will be to establish all Ward identities of the model \`{a} la \cite{Dudal:2021dec}, in presence of the new operators which can be properly introduced via sources. Even in the more trivial sounding $U(1)$ case, new relations between counterterms or Green's functions \cite{Dudal:2021pvw} were recently found when doing so. The Ward identities for the vector operators can also disentangle the physical content of a transverse vector particle, next to the trivial longitudinal parts.

All of the operators we presented here are also gauge invariant, thus they have the potential to be studied with lattice field theory simulations, which has been used as the non-perturbative tool to probe in gauge field theories.

Moreover, recently a BRST invariant formulation of the Refined Gribov-Zwanziger action was obtained for the $SU(N)$ Yang-Mills theory \cite{Capri:2015ixa,Capri:2016aqq,deBrito:2023qfs}. A similar formulation for the $SU(2)\times U(1)$ Higgs model combined with the BRST invariant operators discussed here would allow studying the possible effects of Gribov-Singer copies \cite{Gribov:1977wm,Singer:1978dk} directly at the level of the physical spectrum of the theory. As far as we know, until now, there are not many works on this topic, we can mention \cite{Capri:2013gha,Capri:2013oja} that used the conventional approach to study the $SU(2)\times U(1)$ Higgs model. Such approach would allow to add a non-perturbative ingredient when the coupling grows.

\section*{Acknowledgments}

The authors would like to thank the Brazilian agencies CNPq and FAPERJ for financial support.  S.P.~Sorella is a level $1$ CNPq researcher under the contract 301030/2019-7. G.~Peruzzo is a FAPERJ postdoctoral fellow in the P{\'O}S-DOUTORADO NOTA 10 program under the contracts E-26/205.924/2022 and E-26/205.925/2022.



\begin{thebibliography}{99}

\bibitem{Dudal:2020uwb}
D.~Dudal, D.~M.~van Egmond, M.~S.~Guimaraes, L.~F.~Palhares, G.~Peruzzo and S.~P.~Sorella,
Eur. Phys. J. C \textbf{81}, no.3, 222 (2021)
[arXiv:2008.07813 [hep-th]].
	
\bibitem{Dudal:2021dec}
D.~Dudal, D.~M.~van Egmond, I.~F.~Justo, G.~Peruzzo and S.~P.~Sorella,
Phys. Rev. D \textbf{105}, no.6, 065018 (2022)
[arXiv:2111.11958 [hep-th]].



\bibitem{Nielsen:1975fs}
N.~K.~Nielsen,
Nucl. Phys. B \textbf{101}, 173-188 (1975)

\bibitem{Piguet:1995er}
O.~Piguet and S.~P.~Sorella,
Lect. Notes Phys. Monogr. \textbf{28} (1995), 1-134
doi:10.1007/978-3-540-49192-7

\bibitem{Piguet:1984js}
O.~Piguet and K.~Sibold,
Nucl. Phys. B \textbf{253}, 517-540 (1985)

\bibitem{Dudal:2019aew}
D.~Dudal, D.~M.~van Egmond, M.~S.~Guimar\~aes, O.~Holanda, B.~W.~Mintz, L.~F.~Palhares, G.~Peruzzo and S.~P.~Sorella,
Phys. Rev. D \textbf{100}, no.6, 065009 (2019)
[arXiv:1905.10422 [hep-th]].

\bibitem{Maas:2020kda}
A.~Maas and R.~Sondenheimer,
Phys. Rev. D \textbf{102}, 113001 (2020)
[arXiv:2009.06671 [hep-ph]].

\bibitem{Fradkin:1978dv}
E.~H.~Fradkin and S.~H.~Shenker,
Phys. Rev. D \textbf{19}, 3682-3697 (1979)

\bibitem{Greensite:2008ss}
J.~Greensite and B.~Lucini,
Phys. Rev. D \textbf{78}, 085004 (2008)
[arXiv:0806.2117 [hep-lat]].

\bibitem{Greensite:2021fyi}
J.~Greensite and K.~Matsuyama,
Symmetry \textbf{14}, no.1, 177 (2022)
[arXiv:2112.06421 [hep-lat]].

\bibitem{Frohlich:1981yi}
J.~Frohlich, G.~Morchio and F.~Strocchi,
Nucl. Phys. B \textbf{190}, 553-582 (1981)

\bibitem{Stueckelberg:1938hvi}
E.~C.~G.~Stueckelberg,
Helv. Phys. Acta \textbf{11}, 225-244 (1938)

\bibitem{Stueckelberg:1938zz}
E.~C.~G.~Stueckelberg,
Helv. Phys. Acta \textbf{11}, 299-328 (1938)


\bibitem{tHooft:1980xss}
G.~'t Hooft, C.~Itzykson, A.~Jaffe, H.~Lehmann, P.~K.~Mitter, I.~M.~Singer and R.~Stora,
NATO Sci. Ser. B \textbf{59}, pp.1-438 (1980)

\bibitem{Weinberg:1967tq}
S.~Weinberg,
Phys. Rev. Lett. \textbf{19}, 1264-1266 (1967)


\bibitem{Haag:1992hx}
R.~Haag,
``Local quantum physics: Fields, particles, algebras,'' Springer Science \& Business Media  (2012).

\bibitem{Gribov:1977wm}
V.~N.~Gribov,
Nucl. Phys. B \textbf{139}, 1 (1978)


\bibitem{Singer:1978dk}
I.~M.~Singer,
Commun. Math. Phys. \textbf{60}, 7-12 (1978)


\bibitem{Capri:2015ixa}
M.~A.~L.~Capri, D.~Dudal, D.~Fiorentini, M.~S.~Guimaraes, I.~F.~Justo, A.~D.~Pereira, B.~W.~Mintz, L.~F.~Palhares, R.~F.~Sobreiro and S.~P.~Sorella,
Phys. Rev. D \textbf{92}, no.4, 045039 (2015)
[arXiv:1506.06995 [hep-th]].

\bibitem{Capri:2016aqq}
M.~A.~L.~Capri, D.~Dudal, D.~Fiorentini, M.~S.~Guimaraes, I.~F.~Justo, A.~D.~Pereira, B.~W.~Mintz, L.~F.~Palhares, R.~F.~Sobreiro and S.~P.~Sorella,
Phys. Rev. D \textbf{94}, no.2, 025035 (2016)
[arXiv:1605.02610 [hep-th]].



\bibitem{Capri:2013gha}
M.~A.~L.~Capri, D.~Dudal, M.~S.~Guimaraes, I.~F.~Justo, S.~P.~Sorella and D.~Vercauteren,
Eur. Phys. J. C \textbf{73}, no.10, 2567 (2013)
[arXiv:1305.4155 [hep-th]].



\bibitem{deBrito:2023qfs}
G.~P.~de Brito, P.~De Fabritiis and A.~D.~Pereira,
Phys. Rev. D \textbf{107}  no.11, 114006 (2023)
[arXiv:2302.04827 [hep-th]].

\bibitem{Capri:2017abz}
 M.~A.~L.~Capri, D.~Fiorentini, A.~D.~Pereira and S.~P.~Sorella,
Eur. Phys. J. C \textbf{77}, no.8, 546 (2017)
[arXiv:1703.03264 [hep-th]].

\bibitem{Ikeda:2023aid}
 R.~Ikeda, S.~Kato, K.~I.~Kondo and A.~Shibata,
[arXiv:2308.13430 [hep-lat]].

\bibitem{Dudal:2019pyg}
 D.~Dudal, D.~M.~van Egmond, M.~S.~Guimaraes, O.~Holanda, L.~F.~Palhares, G.~Peruzzo and S.~P.~Sorella,
JHEP \textbf{02}, 188 (2020)
[arXiv:1912.11390 [hep-th]].

\bibitem{Dudal:2021pvw}
 D.~Dudal, G.~Peruzzo and S.~P.~Sorella,
JHEP \textbf{10}, 039 (2021)
[arXiv:2105.11011 [hep-th]].

\bibitem{Binosi:2022ycu}
 D.~Binosi and A.~Quadri,
Phys. Rev. D \textbf{106}, no.6, 065022 (2022)
[arXiv:2206.00894 [hep-th]].

\bibitem{Gambino:1999ai}
P.~Gambino and P.~A.~Grassi,
Phys. Rev. D \textbf{62}, 076002 (2000)
[arXiv:hep-ph/9907254 [hep-ph]].

\bibitem{Jenny:2022atm}
 P.~Jenny, A.~Maas and B.~Riederer,
Phys. Rev. D \textbf{105}, no.11, 114513 (2022)
[arXiv:2204.02756 [hep-lat]].

\bibitem{Greensite:2018mhh}
J.~Greensite and K.~Matsuyama,
Phys. Rev. D \textbf{98}, no.7, 074504 (2018)
[arXiv:1805.00985 [hep-th]].



\bibitem{Krein:1990sf}
 G.~Krein, C.~D.~Roberts and A.~G.~Williams,
Int. J. Mod. Phys. A \textbf{7}, 5607-5624 (1992)

\bibitem{Alkofer:2000wg}
 R.~Alkofer and L.~von Smekal,
Phys. Rept. \textbf{353}, 281 (2001)
[arXiv:hep-ph/0007355 [hep-ph]].

\bibitem{Cucchieri:2004mf}
 A.~Cucchieri, T.~Mendes and A.~R.~Taurines,
Phys. Rev. D \textbf{71}, 051902 (2005)
[arXiv:hep-lat/0406020 [hep-lat]].

\bibitem{Bowman:2007du}
P.~O.~Bowman, U.~M.~Heller, D.~B.~Leinweber, M.~B.~Parappilly, A.~Sternbeck, L.~von Smekal, A.~G.~Williams and J.~b.~Zhang,
Phys. Rev. D \textbf{76}, 094505 (2007)
[arXiv:hep-lat/0703022 [hep-lat]].

\bibitem{Fischer:2008uz}
C.~S.~Fischer, A.~Maas and J.~M.~Pawlowski,
Annals Phys. \textbf{324}, 2408-2437 (2009)
[arXiv:0810.1987 [hep-ph]].


\bibitem{Li:2019hyv}
S.~W.~Li, P.~Lowdon, O.~Oliveira and P.~J.~Silva,
Phys. Lett. B \textbf{803}, 135329 (2020)
[arXiv:1907.10073 [hep-th]].

\bibitem{Binosi:2019ecz}
D.~Binosi and R.~A.~Tripolt,
Phys. Lett. B \textbf{801}, 135171 (2020)
[arXiv:1904.08172 [hep-ph]].

\bibitem{Dudal:2019gvn}
D.~Dudal, O.~Oliveira, M.~Roelfs and P.~Silva,
Nucl. Phys. B \textbf{952}, 114912 (2020)
[arXiv:1901.05348 [hep-lat]].

\bibitem{Falcao:2020vyr}
A.~F.~Falc\~ao, O.~Oliveira and P.~J.~Silva,
Phys. Rev. D \textbf{102}, no.11, 114518 (2020)
[arXiv:2008.02614 [hep-lat]].


\bibitem{Horak:2021syv}
J.~Horak, J.~M.~Pawlowski, J.~Rodr\'\i{}guez-Quintero, J.~Turnwald, J.~M.~Urban, N.~Wink and S.~Zafeiropoulos,
Phys. Rev. D \textbf{105}, no.3, 036014 (2022)
[arXiv:2107.13464 [hep-ph]].

\bibitem{Hayashi:2021nnj}
Y.~Hayashi and K.~I.~Kondo,
Phys. Rev. D \textbf{103}, no.11, L111504 (2021)
[arXiv:2103.14322 [hep-th]].

\bibitem{Hayashi:2021jju}
Y.~Hayashi and K.~I.~Kondo,
Phys. Rev. D \textbf{104}, no.7, 074024 (2021)
[arXiv:2105.07487 [hep-th]].

\bibitem{Siringo:2022dzm}
F.~Siringo and G.~Comitini,
Phys. Rev. D \textbf{107}, no.9, 096001 (2023)
[arXiv:2210.11541 [hep-th]].


\bibitem{Lechien:2022ieg}
T.~Lechien and D.~Dudal,
SciPost Phys. \textbf{13}, no.4, 097 (2022)
[arXiv:2203.03293 [hep-lat]].


\bibitem{Boito:2022rad}
D.~Boito, A.~Cucchieri, C.~Y.~London and T.~Mendes,
JHEP \textbf{02}, 144 (2023)
[arXiv:2210.10490 [hep-lat]].

\bibitem{Lavelle:1995ty}
M.~Lavelle and D.~McMullan,
Phys. Rept. \textbf{279}, 1-65 (1997)
[arXiv:hep-ph/9509344 [hep-ph]].

\bibitem{Capri:2017npq}
M.~A.~L.~Capri, D.~M.~van Egmond, G.~Peruzzo, M.~S.~Guimaraes, O.~Holanda, S.~P.~Sorella, R.~C.~Terin and H.~C.~Toledo,
Annals Phys. \textbf{390}, 214-235 (2018)
[arXiv:1712.04073 [hep-th]].


\bibitem{Aitchison:1983ns}
I.~J.~R.~Aitchison and C.~M.~Fraser,
Annals Phys. \textbf{156}, 1 (1984)

\bibitem{Capri:2018ijg}
M.~A.~L.~Capri, D.~Dudal, M.~S.~Guimaraes, A.~D.~Pereira, B.~W.~Mintz, L.~F.~Palhares and S.~P.~Sorella,
Phys. Lett. B \textbf{781}, 48-54 (2018)
[arXiv:1802.04582 [hep-th]].

\bibitem{Capri:2013oja}
M.~A.~L.~Capri, D.~Dudal, M.~S.~Guimaraes, I.~F.~Justo, S.~P.~Sorella and D.~Vercauteren,
Annals Phys. \textbf{343}, 72-86 (2014)
[arXiv:1309.1402 [hep-th]].

\bibitem{tHooft:1971qjg}
G.~'t Hooft,
Nucl. Phys. B \textbf{35}, 167-188 (1971)
doi:10.1016/0550-3213(71)90139-8

\bibitem{Fujikawa:1972fe}
K.~Fujikawa, B.~W.~Lee and A.~I.~Sanda,
Phys. Rev. D \textbf{6}, 2923-2943 (1972)
doi:10.1103/PhysRevD.6.2923

\bibitem{Maas:2017wzi}
A.~Maas,
Prog. Part. Nucl. Phys. \textbf{106}, 132-209 (2019)
doi:10.1016/j.ppnp.2019.02.003
[arXiv:1712.04721 [hep-ph]].

\bibitem{Elitzur:1975im}
S.~Elitzur,
Phys. Rev. D \textbf{12}, 3978-3982 (1975)
doi:10.1103/PhysRevD.12.3978
\end{thebibliography}
\end{document}